\documentclass[twocolumn,showpacs]{revtex4}
\usepackage{epstopdf}
\usepackage{mathbbol}              
\usepackage{graphics,graphicx,epsfig,ulem} 

\begin{document}

\title{Coherent optical detection of highly excited Rydberg states using electromagnetically induced transparency}
\date{\today}
\author{A. K. Mohapatra, T. R. Jackson and C. S. Adams}
\affiliation{Department of Physics, Durham University,
Rochester Building, South Road, Durham DH1 3LE, England.}

\begin{abstract}              

We observe electromagnetically induced transparency (EIT) on the $5s \rightarrow 5p$ transition in a room temperature rubidium vapour cell  by coupling the $5p$ state to a Rydberg state ($ns$ or $nd$ with $n=26$ to $124$). 
We demonstrate that the narrow line--width of the EIT resonance (2 MHz) allows precise measurement of the $d$ state fine structure splitting, and together with the sensitivity of the Rydberg state to electric fields, we are able to detect transient electric fields produced by the dynamics of charges within the cell. Coherent coupling of Rydberg states via EIT could also be used for cross--phase modulation and photon entanglement.   

\end{abstract}



\maketitle


In comparison to low energy atomic states, highly excited Rydberg states (with principal quantum $n>25$) display rich many--body behavior due to their enhanced two--body interactions \cite{gallagher,mour98}. For example, a gas of ultra--cold Rydberg atoms is found to evolve spontaneously into a plasma \cite{robi00} and back \cite{kili01}. There is considerable interest in the potential to exploit  the strong dipole -- dipole interaction between Rydberg atoms \cite{pillet06} to realise fast quantum gates \cite{jaksch00,lukin01}. The dipole -- dipole interactions also leads to ionization if the ions are not separately confined \cite{li05}. In addition, electromagnetically induced transparency (EIT) involving a Rydberg transition 
could be used to devise a photonic phase gate for optical quantum computing \cite{fried05}. 

Experimentally, Rydberg atoms are detected indirectly via the ions or electrons produced by an ionization pulse \cite{gallagher}. This detection technique provides high efficiency but is destructive and the atom cannot be re--used. For quantum information applications a non--destructive detection of the Rydberg state is preferable. One possibility is  electromagnetically induced transparency (EIT) which is manifest as an absence of absorption or an associated rapid variation in the dispersion allowing dissipation free sensing of the desired atomic resonance. EIT has been widely studied in atomic vapors \cite{eit_review}, frequently using a $\Lambda$ -- scheme where a coherence is induced between two ground states. Alternatively, in the ladder scheme the coherence is induced between a ground state and an excited state via an intermediate state \cite{gea95}. Previously, excited $d$ states with principal quantum number up to $n=8$ have been observed via optical probing of a ladder system \cite{clark01}. 

In our work, we present experimental results on an EIT ladder system involving highly excited Rydberg states with $n=26 - 124$. We show that the non--destructive probing of a Rydberg level opens up an wide range of possible experiments. In particular we focus on two applications: Firstly we exploit the narrow line--width of the EIT spectra to extend measurement of the $nd$ series fine structure splitting up to $n=96$.
Secondly, we illustrate the potential of Rydberg EIT for detecting the dynamics of ions or electrons in the vicinity of the Rydberg excitation region. As ions can be produced by either interactions \cite{li05} or the laser fields used to excite and trap Rydberg atoms \cite{potv06}, knowledge of the role of ions is essential to the success of quantum gate schemes. As we show Rydberg EIT provides an useful diagnostic of the presence of charges.

The energy levels of $^{85}$Rb relevant to this work and the experimental set-up are shown in Fig. 1. The EIT ladder system consists of a weak probe beam resonant with 
$5s~^2S_{1/2}(F=3) \rightarrow 5p~^2P_{3/2}(F')$ transition and an intense coupling beam resonant with the $5p~^2P_{3/2}(F') \rightarrow nd~^2D_{3/2}$ or $5p~^2P_{3/2}(F') \rightarrow nd~^2D_{5/2}$ as indicated in Fig. 1(a).
The $d$ state hyperfine splitting is negligible. 
The probe beam with wavelength $\lambda_{\rm p}=780.24$~ nm, power 1~$\mu$W and beam size 0.4~mm ($1/{\rm e}^2$ radius) propagates through a room temperature rubidium vapor cell of length 75 mm, Fig. 1(b).  The transmission through the cell as the probe beam is scanned is monitored on a photodiode. The probe laser polarization is varied between linear (vertical or horizontal) and circular using appropriate waveplates.
The coupling beam with wavelength $\lambda_{\rm c}=479.2 - 483.9$~ nm is produced by a commercial doubled diode laser system (Toptica TA-SHG).  The coupling beam counter--propagates through the cell with a power up to 200 mW, a spot size of 0.8~mm ($1/{\rm e}^2$ radius), and linear polarization in the vertical direction.
The vapor cell is placed inside a $\mu$-metal shield to reduce the effect of stray magnetic fields.

\begin{figure}[!t]
\begin{center}
\epsfig{file=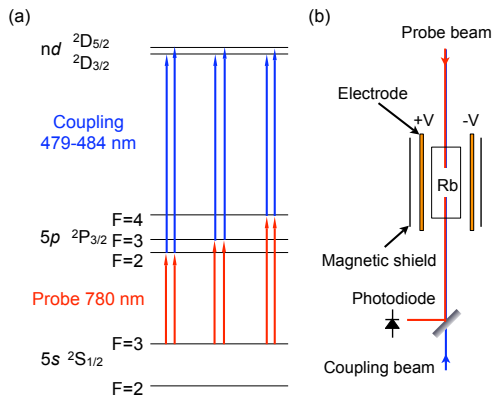,clip=,angle=0,width=7.0cm}
\caption[]{(a) Energy level diagram of the $^{85}$Rb ladder system.
A probe beam at  780 nm measures the absorption on the $5s~^2S_{1/2}\rightarrow 5p~^2P_{3/2}$  transition while an intense coupling beam at 480 nm  dresses the $5p~^2P_{3/2}\rightarrow nd~^2D_{3/2,5/2}$  transition with $n=26 - 124$. The fine and hyperfine splitting of the $nd$ and $5p~^2P_{3/2}$ states respectively give rise to six two--photon resonance lines. The Rydberg state hyperfine splitting is negligible. (b) A schematic of the experimental set-up. The 480 and 780 nm beams counter-propagate through a room temperature Rb vapour cell. The transmission of the 780 nm light is detected on a photodiode. The vapour cell is placed between two electrodes inside a magnetic shield.}
\end{center}
\end{figure}

A typical spectrum  corresponding to the $^{85}$Rb $5s~^2S_{1/2}(F=3) \rightarrow 5p~^2P_{3/2} \rightarrow 45d$ ladder system is shown in Fig. 2. 
In (a) we show the probe absorption with the coupling laser tuned close to and far way from the $5p~^2P_{3/2}(F=4) \rightarrow 45d~^2D_{5/2}$ resonance. The frequency axis is calibrated using the known splittings between the $5p~^2P_{3/2}$ hyperfine states \cite{rapo03}. Note that due to the Doppler mismatch between the probe and coupling lasers, the hyperfine splitting of the $5p~^2P_{3/2}$ state is scaled by a factor of $1-\lambda_{\rm c}/\lambda_{\rm p}$, and the fine structure splitting of the $nd$ state by $\lambda_{\rm c}/\lambda_{\rm p}$.

\begin{figure}[!t]
\begin{center}
\epsfig{file=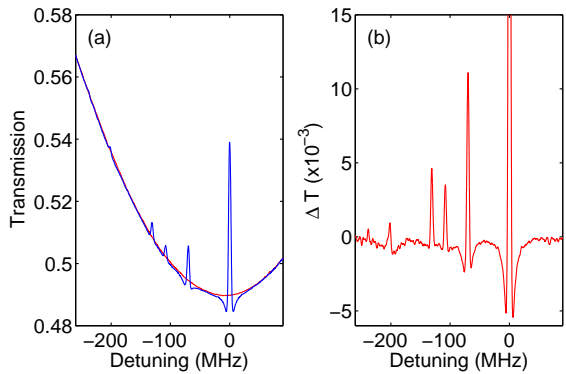,clip=,width=7.5cm}
\caption[]{(a) Probe transmission as a function of probe laser detuning near the $^{85}$Rb $5s~^2S_{1/2}(F=3) \rightarrow 5p~^2P_{3/2}$
resonance with the coupling laser detuned nearly on--resonance (red line)  and far above resonance (blue line)  with the $5p~^2P_{3/2}(F=4)\rightarrow 45d~^2D_{5/2}$ trnasition. (b) The change in the transmission, $\Delta T$, due to the coupling beam given by the difference between the two curves in (a). Six EIT resonances are observed.
}
\end{center}
\end{figure}

In Fig. 2(b) we show the difference between the resonant and far--detuned traces. The spectrum contains six lines corresponding to transitions between the $F=2$, 3, and 4 hyperfine states in the $5p~^2P_{3/2}$ state and both the fine structure components $^2D_{3/2}$ and $^2D_{5/2}$ of the $45d$ state as indicated in Fig. 1(a). The position of the EIT peak within the Doppler broadened absorption profile is determined by the coupling laser detuning. For a coupling power of 180~mW, the largest peak corresponding to the $5p~^2P_{3/2}(F=4)\rightarrow 45d~^2D_{5/2}$ resonance, produces a change in the probe transmission of 5$\%$. This peak height is reduced and the width is increased by about a factor of two if we remove the magnetic shield. The observed spectra are not strongly dependent on the probe laser polarization.

The line--width of the EIT resonance is between $2$ and $4$~MHz depending on the laser power and the transition. The EIT line--width is considerably narrower than the natural width of the intermediate $5p$ states (6~MHz) but much larger than the natural width of the Rydberg states. At low probe laser power, the line--width is limited by the line--width of the probe and coupling lasers and incomplete cancellation of the Doppler shifts due to the wavelength mismatch. The line--width can be broadened to of order 10 MHz by increasing the probe intensity above saturation providing a convenient signal for frequency stabilization of the 480 nm laser.  

Also apparent in Fig.~2(b) is that the lineshape of the EIT feature displays enhanced absorption just below and above the two--photon resonance.
This effect arises due to the wavelength mismatch between the coupling and probe lasers. Although the effect is known \cite{kris05}, it has not been observed in previous experiments. 
One can obtain a theoretical prediction for the EIT lineshape using an approximate expression for the susceptibility derived in the limit of a weak probe \cite{gea95}
\begin{eqnarray}
\chi(v){\rm d}v & = &-i
\frac{3\lambda_{\rm p}^2}{4\pi} \gamma_2 N(v){\rm d}v 
\left[\frac{}{}
\gamma_{2}-i(\Delta_{p}-\mbox{\boldmath$k$}_{\rm p}\cdot \mbox{\boldmath$v$})+\right. \nonumber\\
& ~ &\left.
\frac{(\Omega_{\rm c}/2)^2}{
\gamma_{3}-i(\Delta_{p}+\Delta_{\rm c}
-(\mbox{\boldmath$k$}_{\rm p}+\mbox{\boldmath$k$}_{\rm c})\cdot \mbox{\boldmath$v$})}
\right]^{-1},
\label{eq:chi}
\end{eqnarray}
where $\Omega_{\rm p,c}$, $\Delta_{\rm p,c}$, and $k_{\rm p,c}=2\pi/\lambda_{\rm p,c}$ are the probe or coupling laser Rabi frequencies, detunings and wavevectors, respectively, and $N(v)$ is the number density of $^{85}$Rb atoms with velocity $v$. The decay rates $\gamma_{2,3}$ are the natural widths of the intermediate and upper state in the ladder system. Additional line broadening mechanisms such as laser line--width can be included in $\gamma_3$. By summing the contributions from the different hyperfine lines in the $5p~^2P_{3/2}$ state (with appropriate weightings) and integrating the imaginary part of (\ref{eq:chi}) over the velocity distribution for a room temperature vapor one obtains the absolute absorption coefficient, and hence the transmission through the vapor cell as a function of the probe detuning. Multiple levels in the upper state of the ladder system can be included by adding extra coupling terms in Eq. (1) \cite{badg01}.  
Fig. 3 shows the prediction of Eq. (1) in comparison to the experimental data for (a) $n=45$ and (b) $n=80$. The only fit parameters are the Rabi frequency and detuning of the coupling laser, the rate $\gamma_3$, and the fine structure splitting of the $d$ state.  The enhanced absorption observed at higher $\Omega_{\rm c}$, Fig. 3(a), is accurately predicted by Eq. (1).

\begin{figure}[]
\begin{center}
\epsfig{file=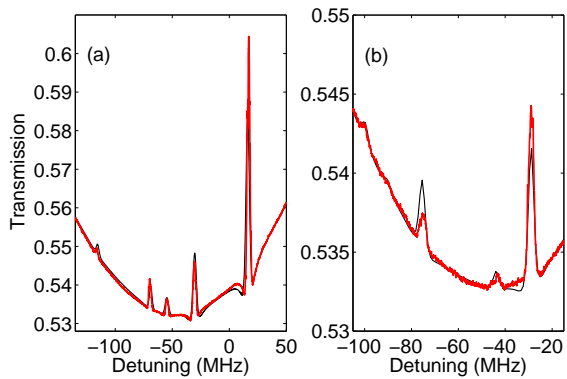,clip=,width=7.5cm}
\caption[]{Probe transmission as a function of detuning (thick red line) compared to Eq. (1) (thin black line) for (a) the $n=45$ $d$ state with $\Omega_{\rm c}= 2\pi(3.5~{\rm MHz})$ and $\gamma_3= 2\pi(0.3~{\rm MHz})$, and (b) $n=80$ with $\Omega_{\rm c}= 2\pi(1.5~{\rm MHz})$ and $\gamma_3= 2\pi(0.3~{\rm MHz})$.}
\end{center}
\end{figure}

For the $80d$ state, Fig. 3(b), the change in the probe transmission is reduced to about 1$\%$. This is consistent with the expected $1/n^{3/2}$ scaling of the coupling beam Rabi frequency. We can still observe the EIT resonance up to the $101d$ state and using lock-in detection up to the $124d$ state. 
The EIT spectra give a direct measure of the $d$ state fine structure splittings. Previous measurements using two--photon absorption and a thermionic diode to detect the ionised Rydberg states have reported values up to $n=65$  \cite{harv77}. We are able to resolve the splitting up to $n=96$. In Fig. 4 we plot the fine structure splitting as a function of $n$ together with data from \cite{harv77}. The line shows a fit assuming a form $A/n^{*3}$ with $A=11.5(4)\times10^3$~GHz and $n^*=n-\delta$, where $\delta=1.35$ is the quantum defect. The $n=75$ spectrum displays an asymmetric lineshape which leads to a systematic error. Our measured splittings are typically larger than those observed in \cite{harv77} by up to 10 MHz.

Finally, we investigated the effect of an external electric field on the Rydberg energy levels by applying a voltage between the copper bar electrodes shown in Fig.~1(b). We observe no effect on the EIT spectra for dc electric fields up to 100 Vcm$^{-1}$ regardless of the probe and coupling laser polarizations suggesting that the Rydberg atoms are being screened. The main source of charge within the cell appears to be ions and electrons produced by photo -- desorption induced by the coupling laser at the surface of the cell \cite{xu96}. For example, by retro--reflecting  the 480 nm laser with a glancing angle at the cell wall we can generate an asymmetric charge distribution which creates a field of order 50~mV/cm in the EIT interaction region. We can detect this small field as a 2.5~MHz splitting of the $70d$ line.

\begin{figure}[]
\begin{center}
\epsfig{file=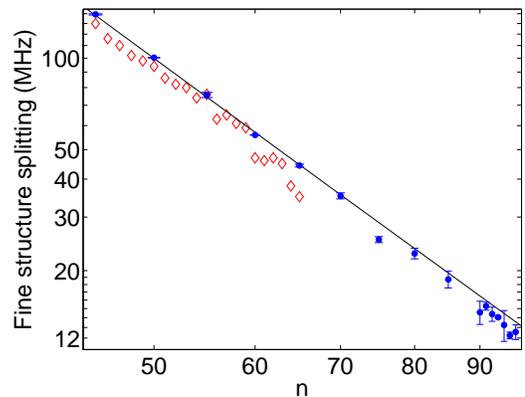,clip=,width=7.0cm}
\caption[]{The observed  fine structure splitting (filled circles) of the $d$ state as a function of the principal quantum number $n$. The open diamonds are data from Ref. \cite{harv77}. The line is a fit of the form $A/n^{*3}$ with $A=11.5(4)\times10^3$~GHz.}
\end{center}
\end{figure}

\begin{figure}[]
\begin{center}
\epsfig{file=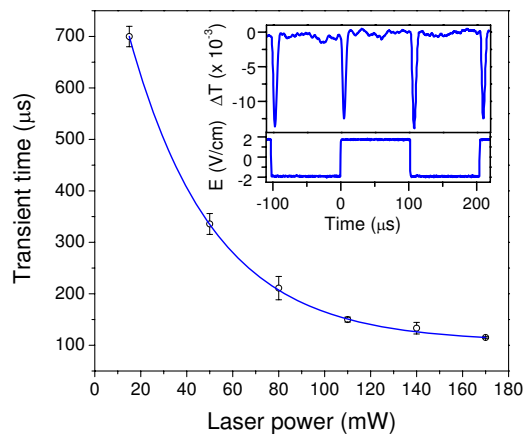,clip=,width=7.0cm}
\caption[]{The width of the EIT transient as the electric field is switched between $\pm 10$~V/cm as a function of the coupling laser power. The transient time decays exponentially to a fixed value at high power. Inset: The transient EIT response as the field is switched between $\pm 2$~V/cm.}
\end{center}
\end{figure}

To further investigate the screening effect and charge dynamics within the cell, we studied the effect switching the electric field direction in a time of order 1~$\mu$s. For a dc field, the charges drift with a distribution which exactly compensates the applied field. However, if the field is switched in a time less than the relaxation time of the charge distribution there is incomplete cancellation of the applied field leading to a perturbation of the EIT signal. 
To monitor the response of the EIT peak to a time varying field we lock the probe laser to the $5s~^2S_{1/2}(F=3) \rightarrow 5p~^2P_{3/2}(F=4)$ transition using polarization spectroscopy \cite{pear02}, and tune the coupling laser to resonance with the $5p~^2P_{3/2}(F=4) \rightarrow 45d~^2D_{5/2}$ transition. The change in transmission ($\Delta T$) as the field is switched is shown in Fig.~5(inset). Each time the field direction changes the EIT peak is suppressed due to the transient penetration of the applied field.
The duration of this transient depends linearly on the amplitude of the field, and decreases exponentially with the coupling laser power, see Fig.~5. The dependence on the coupling laser power arises because at higher power the charge density is increased and the distribution can relax faster towards the steady--state fully--screened distribution.
There is a small asymmetry between the response to increasing and decreasing field that depends on the alignment of the EIT probe relative to the center axis of the cell.  

If we apply an rf frequency to the electrodes then the time variation of the field is too fast for either the ions or electrons to screen the field. In this case, the $^2D_{5/2}$ and  $^2D_{3/2}$ resonances are split into 3 and 2 lines corresponding to their respective $\vert m_J\vert$ components, as shown in Fig. 6.  Similar spectra have recently been observed by ionizing ultra--cold Rydberg atoms \cite{grab06}. In a time varying field, one may not expect to observe narrow lines, however, due to the quadratic dependence on the electric field, the time dependence of the resonance lines has a rectified $\sin^2$--dependence which is almost like a dc field.  The distortion of the lineshape due to the time--dependence of the field is most apparent in the $^2D_{5/2}(\vert m_J\vert=1/2)$ line (at $-120$~MHz in Fig.~6(d)) as it has the highest field sensitivity.
These results indicate that Rydberg EIT could provide a useful optical probe of photo--ionization processes and plasma dynamics. 

\begin{figure}[]
\begin{center}
\epsfig{file=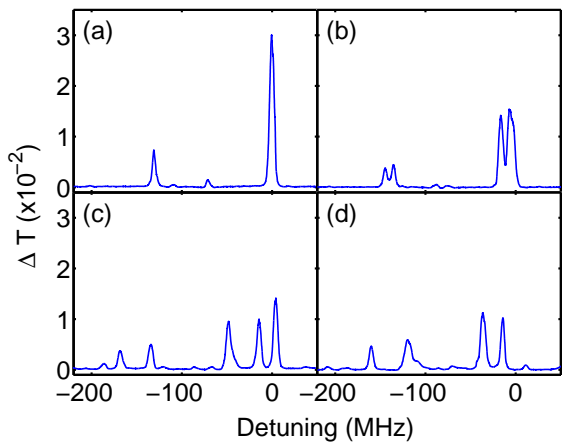,clip=,width=7.5cm}
\caption[]{EIT spectra observed by scanning the coupling laser across the $5p~^2P_{3/2}(F=4) \rightarrow 45d~^2D$ resonance with the probe laser locked to the $5s~^2S_{1/2}(F=3) \rightarrow 5p~^2P_{3/2}(F=4)$ transition. A higher probe power is used so the lines are power broadened and the enhanced absoprtion effect is lost. An rf field with frequency 90 MHz and amplitude (a) 0; (b) 180; (c) 320; and (d) 480 mV/cm is applied. The largest and next largest peak in (a) are the $^2D_{5/2}$ and $^2D_{3/2}$ lines, respectively. As the field is increased these lines are split into 3 and 2 $\vert m_J\vert$ states. In (d) the $^2D_{3/2}(\vert m_J\vert=1/2)$ states have moved out of frame to the left.}
\end{center}
\end{figure}

In summary, we have demonstrated the coherent optical detection of highly excited Rydberg states (up to $n=124$) using EIT, providing a direct non--destructive probe of Rydberg energy levels. The observed spectra, display enhanced absorption below and above the Rydberg resonance, and can be accurately predicted by the optical Bloch equations. The narrow line--width of the EIT resonance allows us to extend measurements of the fine structure splitting of the $nd$ series up to $n=96$. We also show that the EIT spectra are sensitive to charge dynamics within the cell. In our experiment the density of atoms participating in the EIT resonance is more than two orders of magnitude lower than the density used in the observation of dipole blockade \cite{pillet06}.  In future work we will apply the technique to higher density vapors to investigate dipole blockade effects and their potential application in photon entanglement \cite{fried05}.   


We are grateful to S. L. Cornish, I. G. Hughes, M. P. A. Jones and R. M. Potvliege for stimulating discussions and A. P. Monkman for the loan of equipment.
We also thank the EPSRC for financial support.

\end{document}